\documentclass[a4paper,11pt]{article}
\usepackage{jinstpub} % for details on the use of the package, please see the JINST-author-manual
\usepackage{lineno}
\usepackage{hyperref}

\title{\boldmath Picosecond Precision Heavy-Ion Detector for $\Lambda$ Hypernuclei Lifetime Studies}

\author[a]{S.~Zhamkochyan,}
\author[a,1]{S.~Abrahamyan,\note{Corresponding author.}}
\author[a]{A.~Margaryan,}
\author[a]{H.~Elbakyan,}
\author[a]{A.~Kakoyan,}
\author[a]{S.~Mayilyan,}
\author[a]{A.~Papyan,}
\author[a]{H.~Rostomyan,}
\author[a]{A.~Safaryan,}
\author[a]{G.~Sughyan,}
\author[a]{N.~Margaryan,}
\author[a]{G.~Ayvazyan,}
\author[b]{J.~Annand,}
\author[b]{K.~Livingston,}
\author[b]{R.~Montgomery,}
\author[c]{P.~Achenbach,}
\author[c]{J.~Pochodzalla,}
\author[d]{D.L.~Balabanski,}
\author[e]{S.N.~Nakamura,}
\author[f] {A.~Aprahamian,}
\author[a]{V.~Kakoyan}

\affiliation[a]{A.I.~Alikhanyan National Science Laboratory (Yerevan Physics Institute),\\
2 Alikhanyan Brothers., Yerevan, Armenia}
\affiliation[b]{School of Physics \& Astronomy, University of Glasgow,\\
G12 8QQ Scotland, UK}
\affiliation[c]{Institut für Kernphysik, Johannes Gutenberg-Universität Mainz,\\
Mainz, Germany}
\affiliation[d]{Extreme Light Infrastructure- Nuclear Physics (ELI-NP),\\
Bucharest-Magurele, Romania}
\affiliation[e]{Department of Physics, Graduate School of Science, the University of Tokyo,\\
Tokyo, Japan}
\affiliation[f]{Department of Physics and Astronomy, University of Notre Dame,\\
Notre Dame, IN 46556, USA}

\emailAdd{aserg@yerphi.am}

\abstract{In this paper, we present the design and preliminary performance evaluation of a new heavy-ion detector for direct measurements of heavy $\Lambda$ hypernuclei lifetime. The detector employs the previously developed 10 picosecond resolution Radio Frequency (RF) Timer, which converts the temporal information of incident particles into spatial coordinates of secondary or photoelectrons on a position-sensitive detector by means of circular RF scanning in the 500–1000~MHz range. Here, we report the detector design to achieve efficient suppression of accidental background and effective separation of prompt reaction products and delayed events from $\Lambda$ hypernuclei decays, results of test studies carried out with RF synchronized laser as well as preliminary results obtained by using alpha particles. Dedicated Monte-Carlo simulations have been performed to estimate the detector’s performance under realistic experimental conditions at RF-driven electron, photon, or proton beams. The results confirm the feasibility of the proposed design and provide a basis for upcoming experimental measurements, based on the delayed fission detection.}


\begin{document}
\maketitle
\flushbottom

\section{Introduction}
\label{sec:intro}
A free $\Lambda$ particle decays weakly into a nucleon and a pion with a lifetime of $263.2\pm2.0$~ps~\cite{pdg_2012}. The $\Lambda$ lifetime decreases significantly within a hypernucleus due to the availability of additional decay channels, such as the two-body $\Lambda$N$\rightarrow$NN (+176~MeV) (\cite{hypernuclei_nonmesonic_weak_decays,hypernuclei_weak_decays_theory,weak_decays_B_C}) and three-body channels $\Lambda$NN$\rightarrow$nNN (+176~MeV). Since the $\Lambda$ particle can only interact in short range with a restricted number of nucleon pairs (\cite{weak_decays_He_C,decay_width_H_He,weak_decay_H_He}), the hypernuclear lifetime is expected to "saturate" for medium to heavy hypernuclear masses. Two recent comprehensive theoretical calculations of $\Lambda$ hypernuclei lifetimes across the full mass range (up to A=209) have predicted consistent saturation values of 190~ps~\cite{hypernuclei_nonmesonic_weak_decays} and 220~ps~\cite{hypernuclei_weak_decays_theory}. However, experimentally confirming this saturation behavior, especially through precise lifetime measurements of heavy hypernuclei, remains a significant challenge. Relatively accurate measurements of the lifetimes of $^4_\Lambda$H, $^4_\Lambda$He, and $^5_\Lambda$He, $^9_\Lambda$Be, $^{11}_\Lambda$B and $^{12}_\Lambda$C \cite{weak_decays_B_C,weak_decays_He_C,decay_width_H_He,weak_decay_H_He}, employing the emulsion technique, revealed a rapid decrease in lifetime, from a value close to that of a free $\Lambda$ particle to approximately 200~ps in hypernuclei. Measuring the lifetimes of heavier hypernuclei using emulsion techniques is particularly challenging due to the reduced production yields as A increases.

A more advanced experiment at KEK later employed the ($\pi+$, $K+$) reaction, coupled with $\Lambda$-weak decay \cite{lifetime_medium_hypernuclei}, to positively identify hypernuclei from their ground and low-lying excited states, particularly in heavier targets. Using the time-of-flight (TOF) technique and a production reference time, the experiment directly measured decay time spectra for $^{11}_\Lambda$B, $^{12}_\Lambda$C, $^{27}_\Lambda$Al, $^{28}_\Lambda$Si, and $^{56}_\Lambda$Fe. This method, combining precise hypernuclear identification with direct decay time measurement, yielded the most reliable results, showing lifetimes of $211\pm13$~ps, $231\pm15$~ps, $203\pm10$~ps, $206\pm12$~ps  and $215\pm14$~ps respectively~\cite{hypernuclei_mainz_db}. These findings suggest a saturation of hypernuclear lifetime for $A>12$. However, extending this approach to even heavier hypernuclei faces significant challenges due to the dramatically lower production yields for low-lying hypernuclei as A increases further.
Heavy hypernuclear lifetime measurements using the delayed fission technique, combined with the recoil shadow method, were first performed in Kharkov \cite{hypernuclei_kharkov_1986,hypernuclei_kharkov_1987}, where hypernuclei were produced via electron bombardment of bismuth targets, and at CERN through antiproton interactions with bismuth and uranium targets~\cite{hypernuclei_antiproton_cern}. Reported lifetimes for hypernuclei in the bismuth region were $2.7\pm0.5$~ns \cite{hypernuclei_kharkov_1986}, $250\pm100$~ps~\cite{hypernuclei_kharkov_1987} and $180\pm40stat\pm60sys$~ps~\cite{hypernuclei_antiproton_cern}, while those for uranium hypernuclei were significantly shorter, at $130\pm30stat\pm30sys$~ps~\cite{hypernuclei_antiproton_cern}. To clarify these conflicting results, the COSY-13 collaboration conducted subsequent experiments, using a similar method and employing a 1.5~GeV proton beam on gold, bismuth, and uranium targets \cite{hypernuclei_cosy13_2002,hypernuclei_cosy13_1997,hypernuclei_cosy13_2003}. The goal was to increase the momentum transfer to the hypernuclei, thereby enhancing their decay range and reducing measurement uncertainty. The reported fitted lifetime was $145\pm11$~ps, averaged Au, Bi and U. However, these lifetime measurements were indirect, relying heavily on theoretical models to describe the system's time evolution. 

Another lifetime experiment was conducted at JLab \cite{hypernuclei_jlab_2018}, providing a direct measurement of the lifetime of medium-heavy hypernuclei. These hyper-fragments were produced via fission or breakup of heavy hypernuclei initially generated by a 2.34~GeV photon beam incident on thin Fe, Cu, Ag, and Bi target foils. For each event, coincident fragment pairs were detected using a low-pressure multi-wire proportional chamber system \cite{nim_multiwire_1999}. The lifetime was determined from the decay time spectrum, calculated as the time difference between the zero points of the detected fragment pairs. The measured lifetime for each target represents a statistical average over a range of hypernuclear masses, with the mean mass being approximately half that of the target material. The results indicate a lifetime of about 200~ps, with statistical errors ranging from $\pm$3.5~ps to $\pm$49.3~ps and a systematic error of $\pm$10~ps. 

In conclusion, it would be beneficial in view of a more solid comparison with the theoretical predictions, to have new measurements performed at lower energies and by using different experimental techniques. A measurement of the lifetime to a precision of a few percent will guide and constrain the theoretical input, leading to a more precise determination of the hyperon-nucleon interaction.

We present the design and preliminary performance evaluation of a new heavy-ion detector for direct measurements of heavy $\Lambda$ hypernuclei lifetime at RF-driven electron, photon or proton beams. The system integrates the Radio Frequency Timer (RFT) as its core timing device and has been optimized to suppress accidental background and enable efficient separation of prompt and delayed events. 

\section{Heavy-Ion Detector Concept and Design}
\label{sec:detector}
An advanced RFT technique for single electrons was recently developed and tested at AANL \cite{rftimer_nim}. This method utilizes a helical deflector~\cite{helical_nim_2015} that performs circular or elliptical sweeps of keV electrons using a 500–1000~MHz radio frequency electromagnetic field. By translating the arrival time of incident electrons into a hit position on a circle or ellipse, the device achieves exceptionally high ($\sim$10~ps) timing precision. Electron detection is carried out using a position-sensitive detector (PSD) based on microchannel plates (MCP) and a delay line anode.

\begin{figure}[htbp]
\centering
\includegraphics[width=0.7\textwidth]{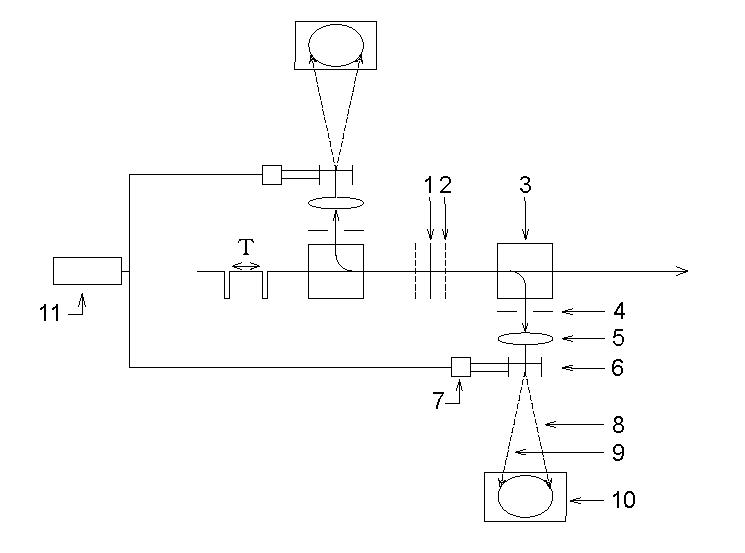}
\caption{The schematic of the Heavy-Ion Detector. 1~-~target, 2~-~accelerating electrode, 3~-~magnet, 4~-~collimator, 5~-~electrostatic lens, 6~-~deflector, 7~-~RF synthesizer, 8,9~-~deflected electrons, 10~-~PSD, 11~-~accelerator's RF system. \label{fig:setup_spectrometer}}
\end{figure}

The Radio Frequency Heavy-Ion Detector (RF HID) presented here builds on this RFT principle for applications in nuclear physics, specifically for direct lifetime measurements of heavy (A$\sim$200) $\Lambda$ hypernuclei. Figure~\ref{fig:setup_spectrometer} illustrates its operational setup. Primary electron, photon, or proton bunches, driven by the RF system of the accelerator, hit the target~(1) and produce secondary electrons~(SE).  Reaction products from beam particles, such as prompt or delayed fission fragments~(FF) can also generate SEs, at a level of tens of SEs per fragment. The SEs emitted from both sides of the target are accelerated by an electrical potential ($\sim$2.5~kV) applied between a cathode (target) and an accelerating electrode~(2). They  are deflected by 90$^\circ$ using a permanent magnet~(3), collimated~(4), focused by an electrostatic lens~(5) and circularly scanned by the deflector~(6) driven by the RF synthesizer~(7) synchronized with the accelerator’s RF system~(11). Coordinates of the deflected electrons are measured by the PSD~(10). 

The PSD, consisting of an MCP array (providing an electron multiplication factor of $10^6$) and a DLD40 delay-line anode~\cite{roentdek}, generates position-sensitive signals with nanosecond-scale rise times. The output signals are processed by a digital acquisition system, consisting of  Picoscope~\cite{picoscope} and a PC. The SEs from the fragments are captured in two opposing arms of the detector with a time resolution of approximately 10-30~ps for coincident events, depending on the aperture size of the SE collimator. This precise timing of both FF minimizes interference from background signals and dark noise in the microchannel plate detectors and improves time resolution by a factor of $\sqrt{2}$. 

Secondary electrons (SEs) detected in the system are originated from two primary sources. The first comprises SEs emitted directly as a result of the primary beam (electrons, photons, or protons) traversing the metallic target. These prompt SEs are generated on both sides of the target surface. The second source involves SEs produced by charged particles resulting from nuclear reactions in the target caused by primary particles. These reaction products, when exiting the target, also emit SEs as they cross its surface. In this case, the times of the produced electrons are determined by the times of nuclear reactions, which are mostly instantaneous.

However, among the nuclear reaction products are also hypernuclei, which are stopped within the target material. These hypernuclei can undergo weak decays after characteristic lifetime of approximately 200~ps, emitting charged particles (pions, protons, fission fragments) that subsequently cross the target surface and produce delayed SEs. These delayed SEs are the key signal for hypernuclear lifetime measurements. Accurate reconstruction of the hypernuclei lifetime requires precise measurement of the time distribution of these delayed SEs.

Another possible outcome to consider is prompt fission of highly excited hypernuclei. In such events, the initial heavy hypernucleus (A$\sim$200) may undergo prompt fission, producing one fragment and one hyperfragment with a typical mass of A$\sim$100. Measurements at Jefferson Lab of proton-, pion-, and kaon-associated photofission of bismuth nuclei in the photon energy range $1.45<E_{\gamma}<1.55$~GeV showed that the fission probability in coincidence with kaons was approximately 16$\%$, about three times larger than proton- and pion-associated fission and roughly twice the inclusive fission probability (see~\cite{bi_photofission_phys_atom_nucl_2010} and references therein). This excess was interpreted as a combination of prompt fission ($\sim7\%$) and formation of bound $\Lambda$ residual states followed by weak nonmesonic decay leading to delayed fission ($\sim9\%$). Hyperfragments with A$\sim$100 can also be formed in kaon-associated prompt fission. However, the fissility of nuclei with A$\sim$100 is approximately three orders of magnitude lower than that of nuclei with A$\sim$200~\cite{fissility_1978}. Consequently, the probability that such hyperfragments undergo delayed fission due to $\Lambda$ weak nonmesonic decay is very small. Therefore, in the proposed experimental configuration, the contribution of delayed fission from A$\sim$100 hyperfragments is expected to be negligible compared to that of the heavy A$\sim$200 hypernuclei of interest.

The main challenge lies in distinguishing delayed secondary electrons (SEs) associated with hypernuclear decay from the overwhelming background of prompt SEs generated directly by the primary beam or by prompt nuclear reactions. In the case of heavy hypernuclei, this separation is significantly simplified. The decay of a $\Lambda$ hyperon in a heavy nucleus induces nuclear fission, producing two fragments emitted nearly back-to-back ($\sim$180$^\circ$). SE yield from each fragment is orders of magnitude larger than the one from a typical primary beam particle. This large SE yield difference and the back-to-back emission geometry allows to use collimation and coincidence requirement between the two detector arms to provide strong suppression of uncorrelated prompt background (see more details in \ref{sec:yields}). 

\begin{figure}[htbp]
\centering
\includegraphics[width=0.55\textwidth]{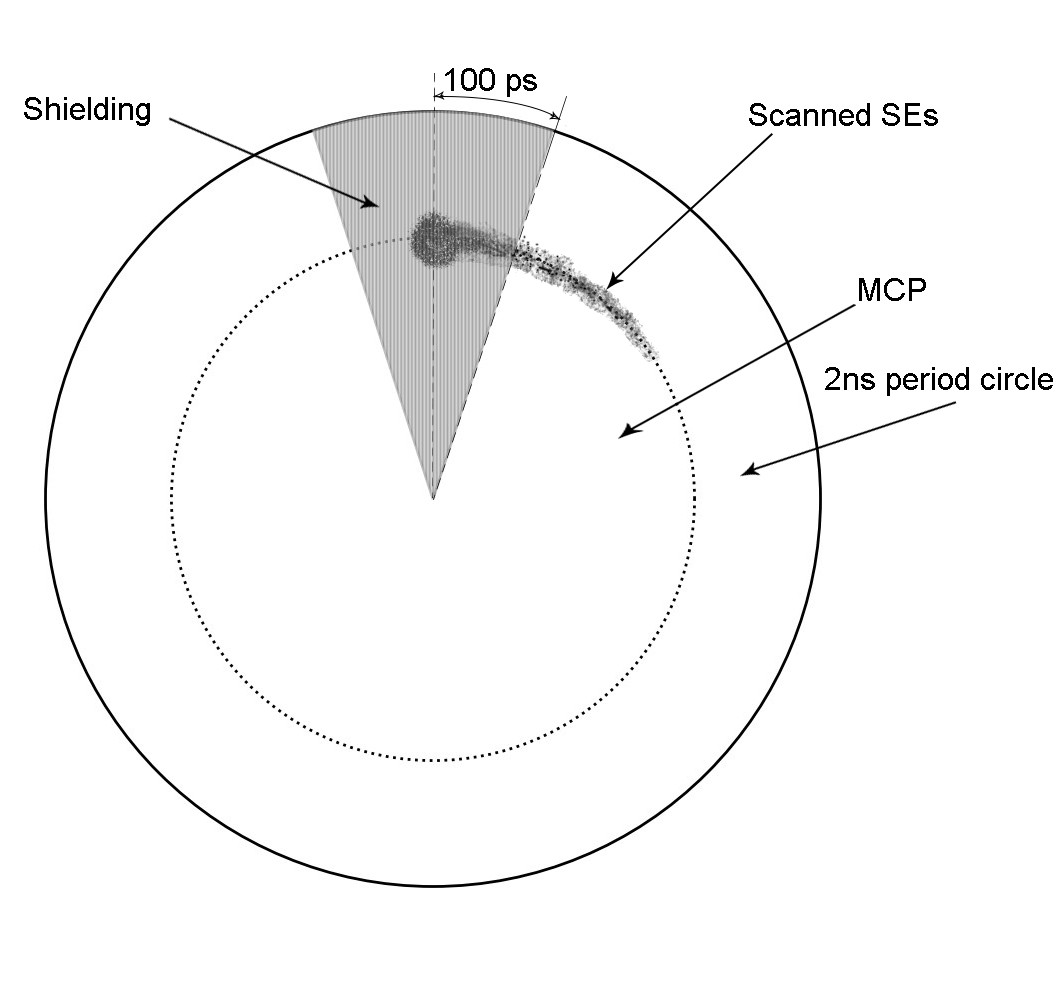}
\caption{Schematic of the position sensitive detection system. \label{fig:pos_sens_det_schema}}
\end{figure}

By selecting an appropriate position-sensitive detector (PSD), the system can be made selectively insensitive to prompt SEs, which arrive at a fixed time with respect to the beam bunch, and hence strike a specific position on the PSD. A $\sim$100~ps length shielding, which corresponds to 18-degree sector (figure~\ref{fig:pos_sens_det_schema}) for the 500~MHz scanning case, could be placed in front of the PSD. This shield will absorb the majority of promptly produced SEs and prevent rapid degradation of the MCPs. The lifetime distribution will then be accurately reconstructed from the detected low rate delayed events that fall outside the shielded phase region. The reference zero-time point can be determined under low beam intensity conditions and subsequently shifted, using a phase shifter, to the center of the shielded sector.

For environments with a substantial radiation background near the beam axis, which could affect PSD and electronics performance, an extended-length RF Timer configuration is proposed. SIMION~8 simulations (section~\ref{sec:montecarlo}) demonstrate that with optimized geometry, secondary electrons can be transported over a 1-meter drift distance without significant transit time spread increase, maintaining sufficient detection efficiency.

\section{Laboratory Tests}
\label{sec:labtests}
\begin{figure}[htbp!]
\centering
\includegraphics[width=0.4\textwidth]{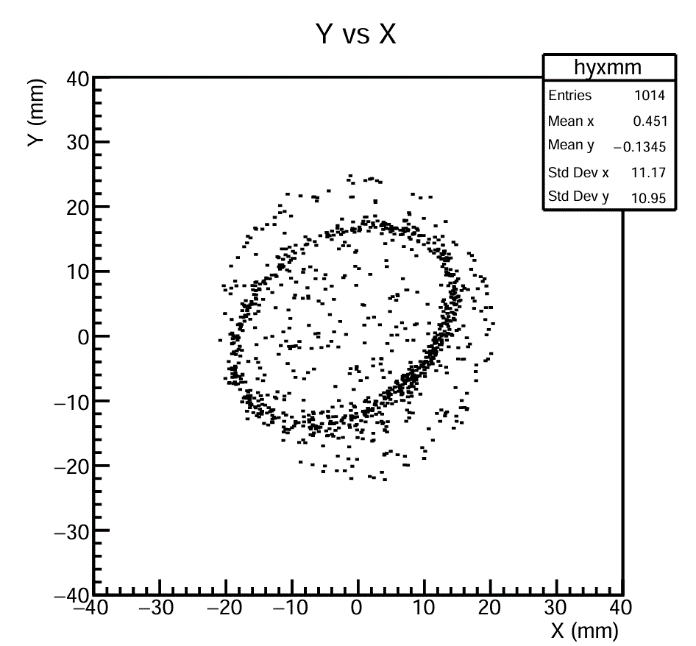}
\caption{Scanned secondary electrons generated by alpha particles from the plutonium source.
\label{fig:alpha_scan}}
\end{figure}

A dedicated setup, employing one arm of the detector, was built in the laboratory for preliminary experimental studies of the detector’s performance. To test the detector’s response to secondary electrons generated by fission fragments, an alpha particle source ($^{239}$Pu) was used in place of the target. The emitted $\sim$5~MeV alpha particles pass through a gold foil and produce SEs, which are then detected by the system. A 1~mm diameter collimator was employed (see Section~\ref{sec:montecarlo} for details on the collimation effects on resolution and efficiency).  The resulting image of the scanned SEs is shown in Fig.~\ref{fig:alpha_scan}. The elliptical image of detected electrons is clearly distinguishable from MCP background noise.

\begin{figure}[htbp]
\centering
\includegraphics[width=0.7\textwidth]{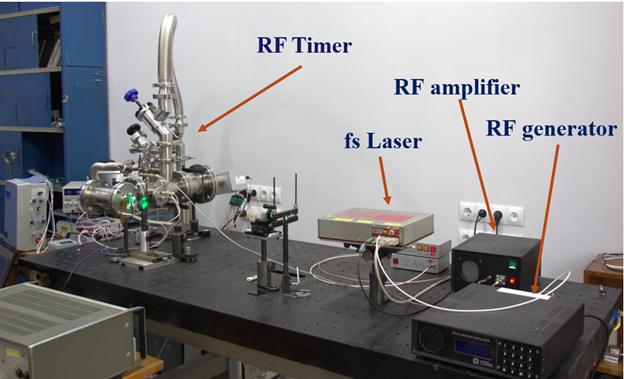}
\caption{Laboratory test setup for RF-synchronized photon measurements.
\label{fig:lab_test_photo}}
\end{figure}

Additional tests were performed using RF-synchronized photons to characterize the system’s time resolution under controlled conditions. These measurements were carried out as part of previous photoemission lifetime studies of 2D materials~\cite{2dmaterials_jinst_2025}. The general layout of the setup is shown in Fig.~\ref{fig:lab_test_photo}. It includes an RFT, a NKT Origami~\cite{nktphotonics} femtosecond laser synchronized with a 500-1000~MHz RF generator, feeding the RF deflector after amplification. The laser delivers 166~fs pulses at a wavelength of 515.4$\pm$4~nm, which is then halved to 257.7~nm using a frequency-doubling crystal. The photon pulses hit the target material and generate photoelectrons, which are subsequently accelerated, deflected and timed by the RFT.

\begin{figure}[htbp!]
\centering
\includegraphics[width=0.85\linewidth]{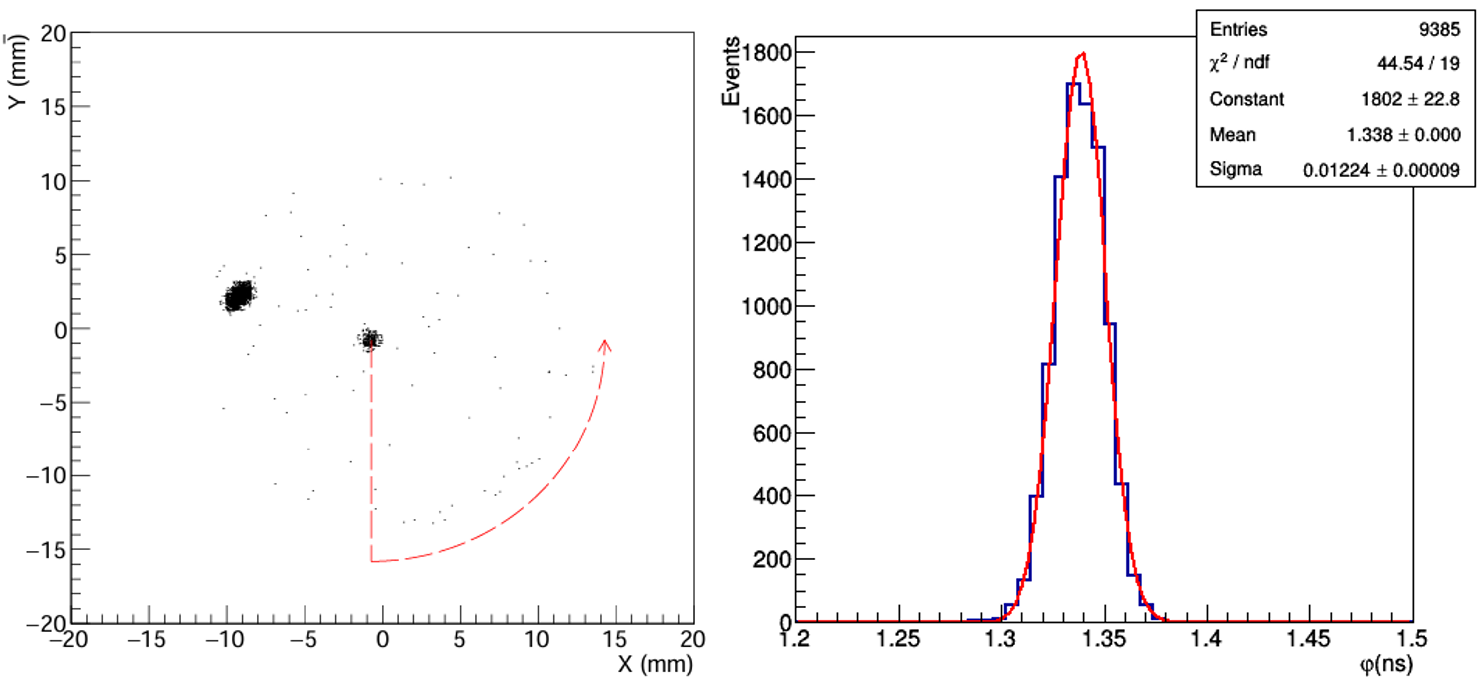}
\par
\vspace{0.3cm}
\includegraphics[width=0.85\linewidth]{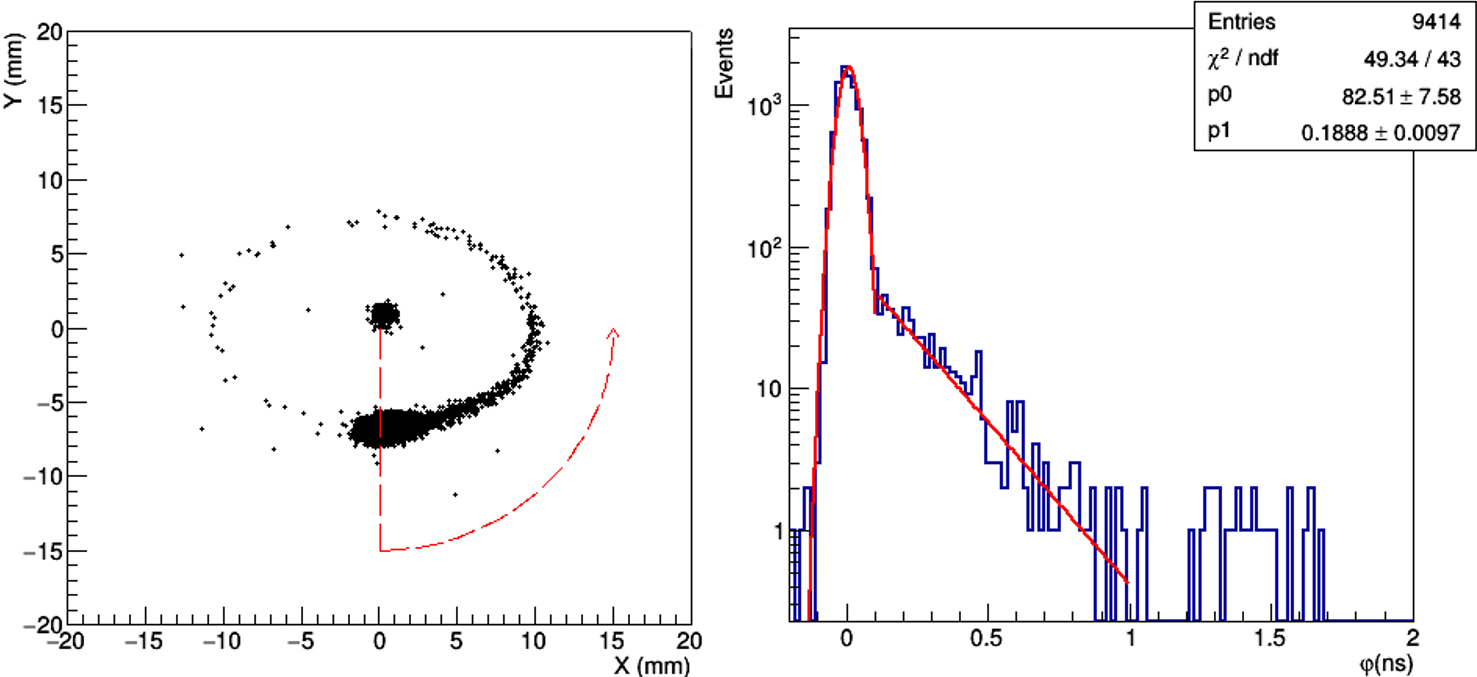}
\caption{Left: Circularly scanned photoelectron image on the PSD. Two runs are overlaid: the RF-Off case shows a central spot used for calibration; the RF-On case produces the circularly scanned image, which can be translated into time information. Right: Corresponding phase distributions in nanoseconds. Top: gold layer; bottom: monolayer graphene.
\label{fig:test_results}}
\end{figure}

Depending on the material under study, the intrinsic photoemission time may vary. Figure~\ref{fig:test_results} (adapted from \cite{2dmaterials_jinst_2025}) (top) shows the hit positions of the circularly scanned photoelectrons (left) and the corresponding phase distribution, converted into time units (right) for a 0.6~µm layer of gold. In this case, the time spread is about 12~ps, which matches the detector’s intrinsic resolution; therefore, these photoelectrons can be considered instantaneous or “prompt”.

Notably, measurements with a monolayer graphene photoemitter revealed a time distribution closely resembling the expected lifetime spectrum of heavy $\Lambda$ hypernuclei decays.  A significant number of prompt photoelectrons were detected, while a smaller part of the events exhibited delays of up to 2 nanoseconds. These extended photoemission times in graphene can be theoretically attributed to its high optical phonon energy and limited phase space for acoustic phonon scattering, which result in hot-carrier lifetimes ranging from hundreds of picoseconds to a few nanoseconds \cite{bistritzer_2009_gr2,lin_fan_2018_gr1}. Experimentally, the nanosecond-scale
photoemission times in graphene at room temperature were first observed using a highly sensitive time-of-flight angle-resolved photoemission spectroscopy (TOF ARPES) technique operated at very low laser fluence ($\sim$10 $\mu$J/cm$^2$) ~\cite{lin_fan_2018_gr1}.

Although in our studies the mean value of the delayed photoemission component varied across graphene samples, an exponential tail was consistently observed. For one of the samples, the result of the fit corresponds to a lifetime of about $189$~ps (figure~\ref{fig:test_results} bottom), comparable to theoretical predictions for hypernuclear lifetime saturation. We fit the prompt part of the graphene spectrum with the Gaussian function and then fit the tail, starting 3 sigma from the Gaussian mean, with an exponential function. The parameter values shown in figure~\ref{fig:test_results} (bottom) correspond to the exponential fit. This result demonstrates the detector’s capability to resolve lifetime spectra with the desired precision.

The essence of the RF timing technique allows the performance of absolute and precise scaling of its time domain, which is a crucial factor for accurate measurement of the lifetime of hypernuclei. Simply, it can be achieved by continuously shifting the phase of the scanning frequency.

\section{Fission Yields and Secondary-Electron Production in the Context of $\Lambda$-hypernuclear Lifetime Experiment Feasibility}
\label{sec:yields}
For the estimation of prompt and delayed event rates, we considered the following configuration: 1.5~GeV electrons delivered in 500~MHz bunches with a duration of 2~ps irradiate a thin bismuth foil of thickness 4~mg/cm$^2$, corresponding to T=0.00064 radiation lengths. At these beam energies and for such a thin target, electron–nucleus interactions are dominated by virtual-photon processes. Electron straggling due to bremsstrahlung can therefore be neglected. The yield Y of a photon-induced reaction channel with cross section $\sigma$(k), where $\sigma$(k) denotes the cross section of the specific process under consideration, can be written in the general form:
\begin{equation}
    \label{eqn:1}
    Y=\frac{NTX_0}{A}\int_{k_{min}}^{E_0}t_{eq}\frac{\sigma(k)}{k}dk,    
\end{equation}
where N is Avogadro’s number, A=209 is the atomic number of bismuth, X$_0$~=~6.3~g/cm$^2$ is the radiation length, and t$_{eq}$=0.017 is the radiation length of the equivalent radiator, which represents the virtual photon flux of the incident electron \cite{tsai_production_leptons}. 

In the 1.5~GeV energy region, lifetime measurements of $\Lambda$-hypernuclei can be performed using a high-energy, high-repetition, photon-induced reactions for which prompt fission cross section is approximately 2.5~mb. Delayed fission is associated with the production of a $\Lambda$-hypernuclei and its subsequent decay, occurring on a timescale of about 200 ps. The dominant production channels are
\begin{equation}
    \label{eqn:2}
    \begin{split}
    Y + p \rightarrow \Lambda + K^+\\
    Y + n \rightarrow \Lambda + K^0.        
    \end{split}
\end{equation}

In a photoproduction experiment on a bismuth target performed at the Thomas Jefferson National Accelerator Facility, correlations between the produced protons, pions, and kaons and the fission of the recoil nucleus were investigated. The cross section for the formation of heavy hypernuclei associated with positive-kaon production (corresponding to only the first reaction in Eq.~\ref{eqn:2}) was estimated to be approximately 12.5~$\mu$b (see~\cite{bi_photofission_phys_atom_nucl_2010} for a detailed discussion). 

Assuming prompt and delayed photo-fission cross sections of 2.5~mb and 12.5~$\mu$b, respectively, in the 1.0–1.5~GeV energy range, 10$^{12}$ electrons traversing a 4~mg/cm$^2$ Bi target are expected to produce, on average, one delayed-fission event and roughly 200 prompt-fission events. Each fission event emits two oppositely directed fragments which, upon exiting the target, generate more than 100 secondary electrons per fragment ($\sim$360 according to~\cite{secondary_electrons_uranium_jnsc}). In contrast, a 1.5~GeV beam electron produces only $\sim$0.03 secondary electrons in the forward direction and $\sim$0.015 in the backward direction \cite{kovalenko_secondary_electrons}. 

For a beam intensity of 10$^{10}$ e/s, each 2~ns period delivers a 2~ps bunch containing $\sim$20 electrons, generating approximately 0.6 forward and 0.3 backward secondary electrons. A beam of this intensity yields about 200 prompt-fission events and at least one delayed-fission event within $\sim$100~s, with each fission event producing more than 100 secondary electrons on each side of the target. The time distribution of these secondary electrons is recorded with the HID system. 

Additionally, by using the PSD design described in section~\ref{sec:detector} (see figure~\ref{fig:pos_sens_det_schema}), the detector can be made effectively insensitive to prompt events, allowing only the low-rate delayed candidate hypernuclear-production events to be registered and processed by the data acquisition system.

Under these conditions, the secondary-electron yield from fission fragments exceeds that from primary beam electrons by more than two orders of magnitude. Therefore, efficient suppression of events induced by beam particles can be achieved through appropriate collimation of the secondary electrons and suitable operating conditions of the detection system. Over 100 hours of beam time, over 3000 delayed-fission events can be accumulated, allowing a determination of the $\Lambda$-hypernuclei lifetime with a statistical precision under 3\%. 

\section{Expected lifetime measurement performance: Monte-Carlo simulations}
\label{sec:montecarlo}
\begin{figure}[htbp!]
\centering
\includegraphics[width=0.95\textwidth]{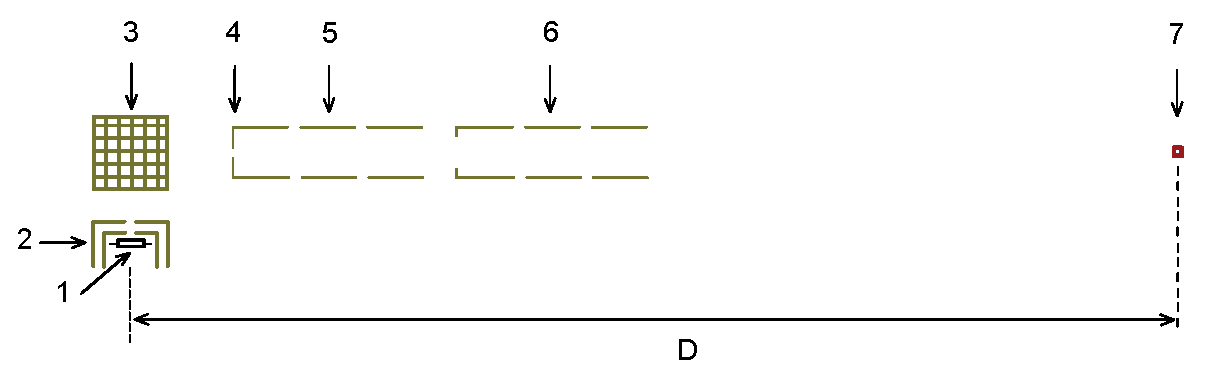}
\caption{The schematic of the simulated detector. 1~-target, 2~-~whenelt-type three-electrodes system, 3~-~magnet, 4~-~collimator, 5,6~-~Einzel lenses, 7~-~detection plane. 
\label{fig:simulation_schema}}
\end{figure}

Simulations based on the SIMION~8 software package~\cite{simion} were conducted to evaluate the detector’s performance under varying geometric configurations and applied static voltages, resulting in different combinations of detection efficiency, focusing distance, and time resolution. The schematic of the simulated setup is shown in Fig~\ref{fig:simulation_schema}. Secondary electrons (SEs) are generated with uniform spatial and angular distributions at emission area on target (1). The size of the emission area corresponds to the cross-section of the high-energy beam hitting a target. The initial energies of secondary electrons emitted by fission fragments have been experimentally measured to lie in the range of 0.3 to 5~eV~\cite{secondary_electrons_alpha_jetp, secondary_electrons_uranium_jnsc}. 

\begin{figure}[htbp]
\centering
\includegraphics[width=0.75\textwidth]{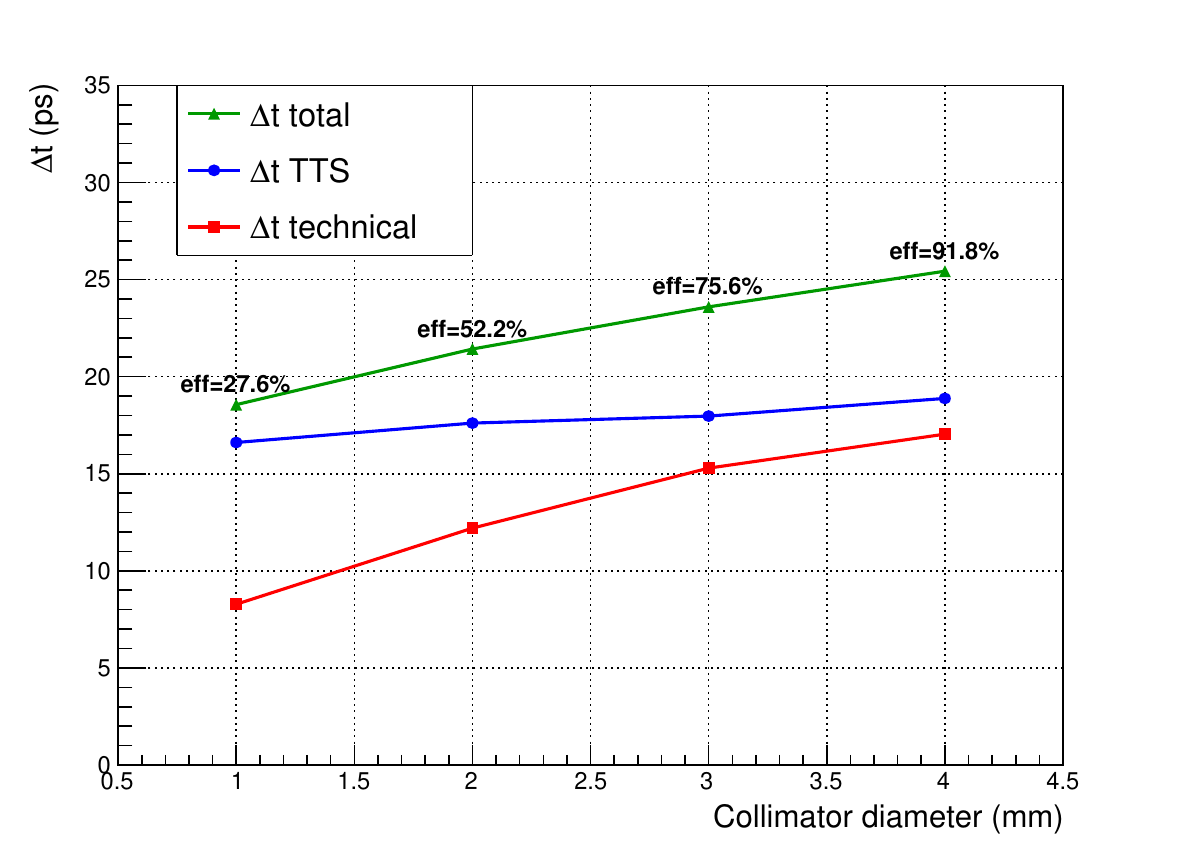}
\caption{Calculated transit time spread, technical time resolution, total time resolution, and SE detection efficiency as functions of collimator diameter for an SE emitter size of 0.5~mm (D~=~50~cm).
\label{fig:times_vs_collimator_size}}
\end{figure}

In the present simulations, the SE energy distribution is taken as exponential, with a mean value of 2~eV. The electrons are pre-focused and accelerated to 2500 eV in a Wehnelt-type three-electrode system (2), deflected by 90~degrees in a permanent magnet (3), and then pass through the focusing system composed of a collimator (4) and two Einzel lenses (5) and (6), arranged sequentially. The transit time spread (TTS) and electron beam spot size at the detection plane (7) are calculated. The TTS represents the detector’s physical time resolution, while the spot size contributes to the technical time resolution, given by: 
\[
\Delta t_{\text{tech}} = \frac{T\Delta d}{2 \pi r},
\]
where T is the period of the RF cycle, $\Delta$d is the convolution of the SE spot size at the detection plane and the position resolution of the PSD, and r is the radius of the scanned circle.  In the simulations we consider T = 2~ns (corresponding to a scanning frequency of 500 MHz), $\Delta$d is taken as the SE spot size at the detection plane (the spot size is in the hundreds of micron range, which is significantly larger than the position resolution of the PSD) and r = 10~mm (typical scan radius for 25~mm diameter MCPs).
Figure~\ref{fig:times_vs_collimator_size} shows the calculated components of time resolution: SE transit time spread ($\Delta t_{tts}$), technical time resolution ($\Delta t_{tech}$), and total time resolution ($\Delta t_{total}$), given by:
\[
\Delta t_{\text{total}} = \sqrt{ \left( \Delta t_{\text{tts}} \right)^2 + \left( \Delta t_{\text{tech}} \right)^2 }
\]

The SE detection efficiency (eff), defined as the ratio of the number of electrons reaching the detection plane to the number generated at the target, is also shown. These parameters are presented as functions of the collimator diameter for an SE emitter size of 0.5~mm and a detection distance of D~=~50~cm from the target.

\begin{figure}[htbp]
\centering
\includegraphics[width=0.75\textwidth]{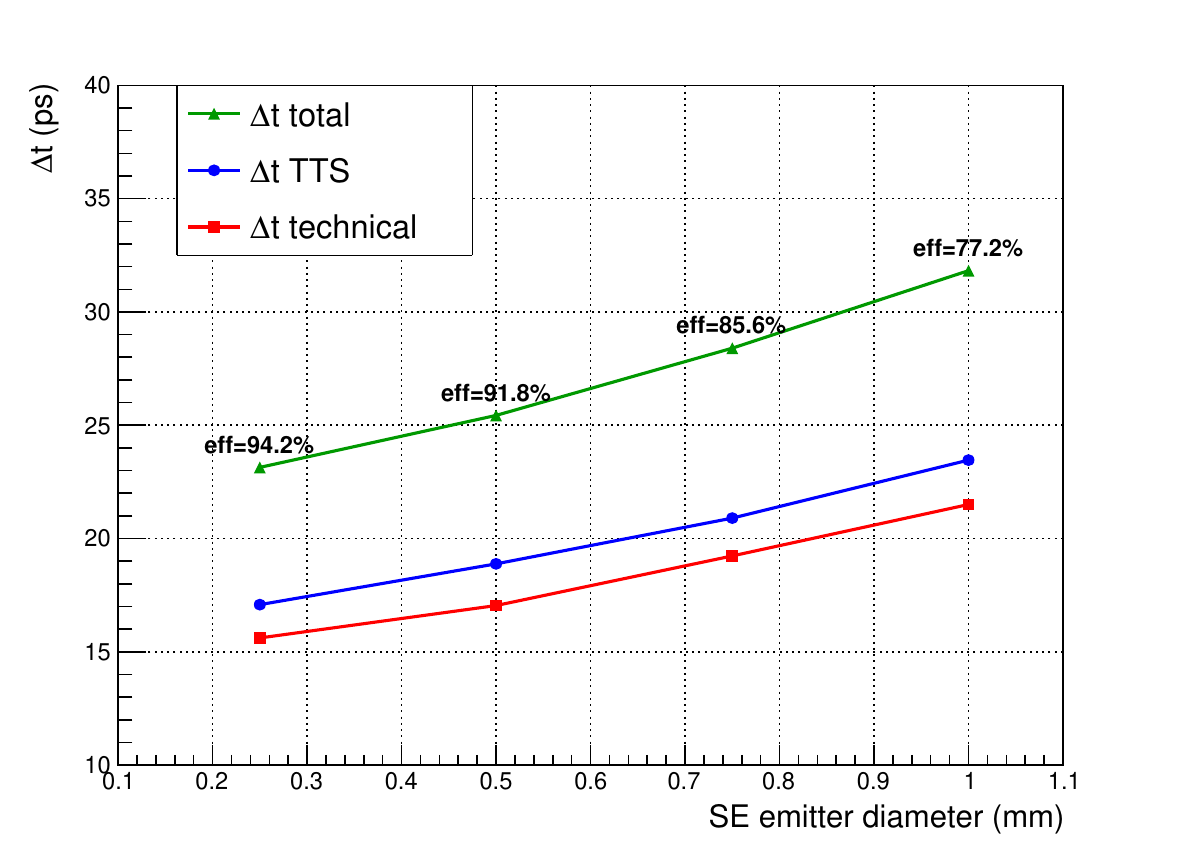}
\caption{Calculated transit time spread, technical time resolution, total time resolution, and SE detection efficiency as functions of SE emitter size for a collimator diameter of 4~mm. 
\label{fig:times_vs_emitter_size}}
\end{figure}

Figure~\ref{fig:times_vs_emitter_size} shows the same parameters calculated for a fixed collimator diameter of 4~mm, but as a function of the SE emitter size. This illustrates the impact of varying the high-energy beam cross-section on the achievable time resolution and detection efficiency.

By varying the collimator aperture we can choose an optimal efficiency and detector resolutions. These calculations show that even in the worst scenario we can achieve $\sim$30ps resolution. Simulations also demonstrate that the secondary-electron transport system can be positioned at distances up to 1~m or more from the target, if needed to mitigate radiation damage to the PSD and associated electronics, without significant degradation of time resolution or detection efficiency.

As discussed in Section~\ref{sec:yields}, the ratio of prompt to delayed fission events is estimated to be 200:1 for a bismuth target, while prompt secondary electrons generated directly by the primary beam can be suppressed through collimation in the detector system. It was also shown that under certain conditions, approximately 3000 delayed-fission events can be collected over 100 hours of beam time. In general, the number of detected delayed-fission events ($N_{delayed}$) and the prompt-to-delayed ratio ($N_{prompt}/N_{delayed}$) will vary depending on the specific experimental conditions: the type and properties of the initial beam, the target and the actual beam time.

For the simulation estimates presented below, we assume $N_{delayed}=1000$ and prompt-to-delayed ratios 10$^3$ and 10$^5$. The measured prompt events times follow a Gaussian distribution around the time zero point with a standard deviation determined by the detector’s time resolution, while the delayed events follow an exponential distribution convoluted with Gaussian noise (again with standard deviation determined by the detector’s time resolution).

\begin{figure}[htbp!]
\centering
\includegraphics[width=1.0\textwidth]{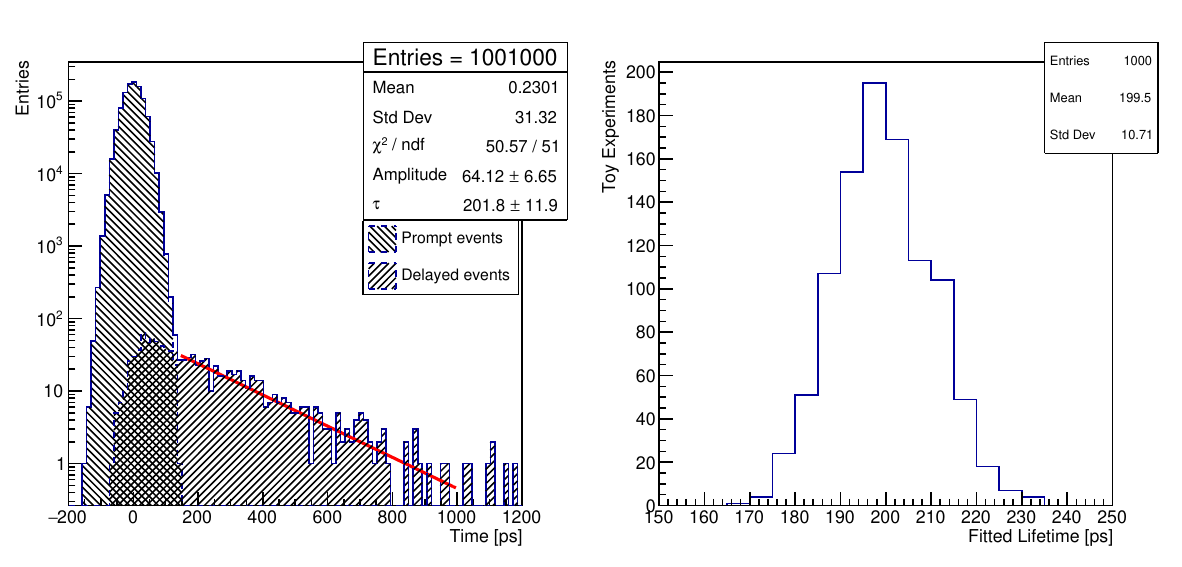}
\caption{Left: Simulated “toy” experiment with $N_{delayed}=1000$ and $N_{prompt}/N_{delayed}=10^3$, detector resolution of $\sigma$=30~ps and "true" lifetime set at 200~ps. The superposition distribution of prompt and delayed events is fitted with an exponential tail in 142.6~ps~<~t~<~1000~ps range. Right: Distribution of fitted lifetimes over 1000 “toy” experiments.
\label{fig:toy_experiment_1000_1000}}
\end{figure}

Our goal is to separate the delayed (exponential) distribution from the prompt distribution, fit it and obtain the main value, which will correspond to the delayed events lifetime. The most straightforward way is to apply a cut for exponential fitting so that prompt Gaussian events will not significantly contaminate the delayed exponential distribution’s remainder after the cut. If we want to expect 1 or fewer prompt events for $t~>~t_{cut}$ then: 
\[
t_{cut} = \sigma \times \Phi^{-1}\!\left(1 - \frac{1}{N_{prompt}}\right),
\]
where $\sigma$ is the detector time resolution and $\Phi^{-1}$ is the inverse of the standard normal cumulative distribution function (the quantile function). 

Figure~\ref{fig:toy_experiment_1000_1000} (left) illustrates an example result of simulated “toy” experiment (a superposition of prompt and delayed events) with $N_{delayed}=1000$ and an event ratio of $N_{prompt}/N_{delayed}=10^3$ and a detector resolution of $\sigma$=30~ps ($t_{cut}=4.7\cdot\sigma=142.6$~ps).  The "true" lifetime was set at 200~ps. The distribution is fitted with an exponential curve in the range $t_{cut}<t<1000$~ps. Figure~\ref{fig:toy_experiment_1000_1000} (right) shows the distribution of lifetimes reconstructed from 1000 such “toy” experiments resulting in an extracted lifetime of $\tau=199.5$~ps and an expected uncertainty of $10.7$~ps. Using the same parameters and $N_{prompt}/N_{delayed} = 10^5$ yields an extracted lifetime of 200.5 ps and an expected uncertainty of 13.5 ps.

\section{Summary}
\label{sec:summary}
A new time-resolved heavy-ion detector, based on circular radio-frequency scanning of secondary electrons, has been designed, built and tested in the laboratory using a synchronized femtosecond laser. The measured time resolution of the detector is approximately 12~ps. This system can be used for the heavy-ion precise timing and is particularly suitable for heavy hypernuclear lifetime studies, enabling the separation of high-rate prompt reaction products from delayed events of interest. Laboratory studies of hot-carrier lifetime in graphene, which exhibit a time structure similar to that expected for hypernuclear decays, demonstrate the detector’s capability for such measurements. Additionally, the detection of secondary electrons generated by alpha particles has been successfully verified. Detailed Monte Carlo simulations confirm that the expected statistical uncertainty for direct hypernuclear lifetime measurements with this detector is about 10~ps under realistic conditions. These results indicate that the proposed detector concept is feasible and provides a solid basis for future experiments with electron, photon, or proton beams.

\acknowledgments
This work was supported by the Higher Education and Science Committee of the Republic of Armenia (Research Project: 23LCG-1C018) and the International Science and Technology Center (ISTC project AM-2803).

\bibliographystyle{JHEP}
\bibliography{biblio.bib}
\end{document}